\documentclass[aps,pre,preprint]{revtex4-1}
\usepackage{url} 
\usepackage[utf8]{inputenc}
\usepackage{epsfig,amssymb,graphicx,subfigure}
\usepackage{amsmath}

\usepackage{hyperref}

\begin{document}

\title{Stochastic Dynamics of Virus Capsid Formation:\\
Direct versus Hierarchical Self-Assembly}
\author{Johanna E Baschek}
\thanks{These authors contributed equally.}
\author{Heinrich C R Klein}
\thanks{These authors contributed equally.}
\author{Ulrich S Schwarz}
\affiliation{Institute for Theoretical Physics and BioQuant, Heidelberg University, Heidelberg, Germany}

\begin{abstract}
In order to replicate within their cellular host,
many viruses have developed self-assembly strategies for their capsids
which are sufficiently robust as to be reconstituted \textit{in
  vitro}. Mathematical models for virus self-assembly usually assume
that the bonds leading to cluster formation have constant reactivity
over the time course of assembly (\textit{direct assembly}). In some
cases, however, binding sites between the capsomers have been reported
to be activated during the self-assembly process (\textit{hierarchical
  assembly}). In order to study possible advantages of such
hierarchical schemes for icosahedral virus capsid assembly, we use
Brownian dynamics simulations of a patchy particle model that allows
us to switch binding sites on and off during assembly. For T1 viruses,
we implement a hierarchical assembly scheme where inter-capsomer bonds
become active only if a complete pentamer has been assembled. We find
direct assembly to be favorable for reversible bonds allowing for
repeated structural reorganizations, while hierarchical assembly is
favorable for strong bonds with small dissociation rate, as this
situation is less prone to kinetic trapping. However, at the same time
it is more vulnerable to monomer starvation during the final
phase. Increasing the number of initial monomers does have only a weak
effect on these general features. The differences between the two
assembly schemes become more pronounced for more complex virus
geometries, as shown here for T3 viruses, which assemble through
homogeneous pentamers and heterogeneous hexamers in the hierarchical
scheme. In order to complement the simulations for this more
complicated case, we introduce a master equation approach that agrees
well with the simulation results.
Our analysis shows for which molecular
parameters hierarchical assembly schemes can outperform direct ones
and suggests that viruses with high bond stability might prefer
hierarchical assembly schemes. These insights increase our physical
understanding of an essential biological process, with many
interesting potential applications in medicine and materials science. 
\end{abstract}

\maketitle

\section{Background}

The structure and dynamics of viruses are a fascinating research
subject not only from a biological, but also from a physical
perspective \cite{Roos2010}. In particular, they are a very
instructive model system to study self-assembly of large protein
complexes with a relatively clear biological function. As viruses do
not show metabolic activity of their own, they need to infect host
organisms in order to replicate.  One key step during the replication
process is the formation of the protein shell containing the viral
genome. For many viruses, the capsid formation is sufficiently
autonomous that it occurs even \textit{in vitro}
\cite{Fraenkel-Conrat1955}.  This robustness of the process guarantees
successful replication within the dynamic and heterogeneous
environment of a living cell.  Although virus shell formation is
considered as a paradigm for the self-assembly of protein complexes
\cite{Johnson1997}, its underlying principles are far from being fully
understood.  Progress in our understanding of virus assembly
would increase our knowledge of a process of large biological and
medical relevance as well as help to advance new self-assembly
strategies in material science applications \cite{Pawar2010}.

A large variety of mathematical models and simulation approaches has
been developed to gain insight into the dynamics of capsid formation
from a theoretical perspective.  In these approaches the
characteristics of protein association and dissociation processes were
analyzed depending on parameters like interaction strength, subunit
geometry or temperature. The employed techniques range from
large-scale Molecular Dynamics (MD) simulations with only a modest
amount of coarse-graining of the atomic details
\cite{Arkhipov2006} through various schemes of coarse-grained MD
\cite{Rapaport2004,Hagan2006,Nguyen2007,Rapaport2008,Rapaport2010,Johnston2010}
to patchy particle simulations with interaction potentials
\cite{Wilber2009,Wilber2009a}. A thermodynamic framework for assembly
of icosahedral viruses has been established by Zlotnick and coworkers
\cite{Moisant2010,Zlotnick1994,Zlotnick1999,Zlotnick2002,Endres2002,Johnson2005,Zlotnick2010}.
In general, these studies have revealed that the formation of complete
virus capsids requires intermediate bond stability. If interaction
strength is too high (or, equivalently, temperature too low), the
system becomes kinetically trapped in intermediates which cannot
reconstruct anymore due to the strong binding. If interaction strength
is too low (or, equivalently, temperature too high), the target
structure is not sufficiently stable. Another mechanism which can
prevent complete capsid formation is the occurrence of misfits, leading
to structural polymorphism as often studied with MD-schemes allowing
for cluster distortions \cite{Nguyen2008,Elrad2008,Nguyen2009}.

Due to the large number of single building blocks assembling during
virus formation (the simplest icosahedral capsid, T1, has already 60
protein components), there is a multitude of topologically possible
assembly pathways.  Similar to protein folding, the dominance of few
key structures is believed to limit the number of pathways and to
speed up the process \cite{Moisant2010}. In this respect it has been
observed that some viruses have developed mechanisms to orchestrate
self-assembly by regulating the reactivity of their binding sites
\cite{Cardarelli2011,Dokland2000}.  This switching establishes a
hierarchy in the formation of transient intermediates during the
assembly process. In a number of experiments, partly supported by
theoretical calculations, it has been shown that intermediates of
pentameric and hexameric symmetry are of special importance for the
assembly process of icosahedral viruses
\cite{Johnson1997,Tonegawa1970,Salunke1986,Flasinski1997,Zlotnick2000,Willits2003,Hanslip2006,Oppenheim2008}.
Early observations of \textit{in vitro} assembly of phages and small
viruses revealed pentamer sub-structures to play a key role
\cite{Tonegawa1970,Salunke1986}.  Experiments on Brome Mosaic Virus
\cite{Flasinski1997}, Cowpea Chlorotic Mottle Virus
\cite{Willits2003}, Human Papillomavirus \cite{Hanslip2006} and Simian
Virus 40 \textit{in vivo} and \textit{in vitro} \cite{Oppenheim2008}
explicitly treat capsid assembly from pentameric capsomers.  A model
for the assembly of Cowpea Chlorotic Mottle Virus suggests that its
protein shell assembles from pentamers as well as from trimers of
dimers (hexamers) \cite{Johnson1997}.

Despite the described variety of computational approaches used for
virus assembly, to our knowledge the effect of a state-dependent activation
of binding sites during the assembly process (\textit{hierarchical
  assembly}) has not been explored yet from the theoretical point of
view.  Although some models consider assembly from pentameric and
hexameric clusters, these subunits at the same time represent the
smallest entities of the system and their formation from single
proteins is not included \cite{Nguyen2008,Wales2009,Johnston2010}.
Here we investigate the effect of a binding hierarchy on the assembly
of icosahedral viruses by comparison of hierarchical and
non-hierarchical (direct) assembly from single monomers.  We use
Brownian Dynamics simulations with reaction patches which have
previously been used to study transport-limited protein reactions
\cite{Schluttig2008,Schluttig2010}.  Our approach assumes well-defined
capsid geometries (in the spirit of local rules) and does not require
the use of interaction potentials.  This makes our simulations
relatively fast, but does not allow us to study structural polymorphism.
One particular strength of our approach is that it implements the correct mobility
matrix for each possible geometry of the assembling clusters
\cite{Schluttig2008}. Another advantage which is exploited here is
that one can easily implement hierarchical assembly by an event-driven
switching of patch reactivity.

This paper is organized as follows. We first give an overview of the
simulation framework and the implemented geometries. Then we present
our results for direct and hierarchical assembly of T1 virus capsids.
The analysis of T1-assembly is completed with a comparison of
the two assembly mechanisms and a discussion of the effect of an
increased number of initial monomers.  We then explain our results for
direct and hierarchical assembly of the more complex T3 virus. They
are followed by a detailed analysis of the formation of individual
capsomers, which includes a master equation approach. The paper closes
with concluding remarks and an outlook to potential future
applications of our approach.

\section{Methods}

\subsection{Outline of the Computer Simulations}

To study virus assembly we use a Brownian Dynamics approach with
patchy particles which has been developed before to investigate
diffusion and association of model proteins and their complexes
\cite{Schluttig2008,Schluttig2010}. Single proteins are modeled as
hard, spherical particles with equal radius. They are equipped with a
specific number of reaction patches representing the binding
sites. The geometry of the virus capsid is coded in the position
of the reaction patches on the spheres.  Assemblies of several proteins are
treated as rigid objects whose diffusive characteristics are
calculated on the fly upon formation \cite{Carrasco1999}.  In each
simulation step the particles are propagated according to their
translational and rotational diffusive properties, followed by
possible association and dissociation steps.  Binding of two proteins
is implemented as a two-step process following the notion of the
encounter complex \cite{Schreiber2009}. Upon diffusional overlap of
two reaction patches, binding occurs stochastically with a predefined
patch-specific rate $k_a$. Thus, the probability for the transition
from encounter to a bound state within a given timestep $\Delta t$ is
$p_\text{bind}=k_a \Delta t$.  If the bond formation is accepted, the
binding partners instantaneously click into their predefined relative
orientation, assuming that his processes is much faster and less
stochastic than diffusion and association. The repositioning is
distributed among the two clusters according to their diffusive
weights. If this reorientation leads to a steric overlap of the two
associating partners with each other or with other protein complexes,
binding is rejected and the old positions and orientations of the
clusters are used for the next simulation step. Similarly to
association, dissociation of an existing bond occurs stochastically
with the bond specific rate $k_d$. Thus, a bond is disrupted within
$\Delta t$ with the probability $p_\text{break}=k_d \Delta t$.  If the
broken bond was the only connection between two clusters, both are
propagated independently in the following simulation step.  The
simulation algorithm is combined with a visualization routine which
enables us to follow the assembly process. Some representative snapshots of the step-wise
assembly of a T1 virus capsid are shown in Figure 1. While the upper
row shows direct assembly from 60 monomers (dark blue), in the lower
row monomers (light blue) first have to form pentamers (red), which
then in turn form the complete capsid (hierarchical assembly).

\subsection{Capsid Geometries}

The capsid geometries follow the well-established Caspar-Klug scheme
where the quasi-equivalent positions in the scaffold of an icosahedral
capsid are represented by different types of monomers
\cite{CasparD.L.DAndKlug1962}.  The structural complexity is described
by the triangulation number T derived from the capsid geometry. T is
restricted to certain integer values (T=1,3,4,7,9,...) and denotes the
number of protein types which are needed to form a full icosahedral
shell. The total number of monomers per capsid is $n_f$=60 T. The
icosahedron vertices represent points around which the proteins
cluster into close-packed arrays. The proteins grouped around the
twelve vertices (which represent axes of fivefold symmetry) form
pentamers, while the triangular faces of the icosahedron are covered
with hexamers. Every scaffold consists of 12 pentameric and $10 \cdot
($T$-1)$ hexameric capsomers. In our description we restrict the
effect of growing complexity (T$>1$) to the hexamers. Thus, every
hexamer contains (T$-1$) different proteins, so that the number of
hexameric subunits as well as the number of individual components of
each hexamer increase with T.  We want to point out that this scheme
for virus geometries into ringlike subunits represents only one out of
several possible realizations.  Following previous approaches to the
characterization of icosahedral geometries \cite{Schwartz1998}, we use
a set of local rules to define the bond angles between the individual
particles. In this way, the resulting structure is encoded in the bond
properties of the elementary subunits. Due to the high symmetry of the
viral capsid, only a small number of different bonds is sufficient to
define a unique target geometry. We note that the exact definition of
bond properties impedes the formation of aberrant cluster
structures. Therefore the approach used here does not allow us to
study structural polymorphism.

\begin{figure}[t]
\includegraphics[width=\textwidth]{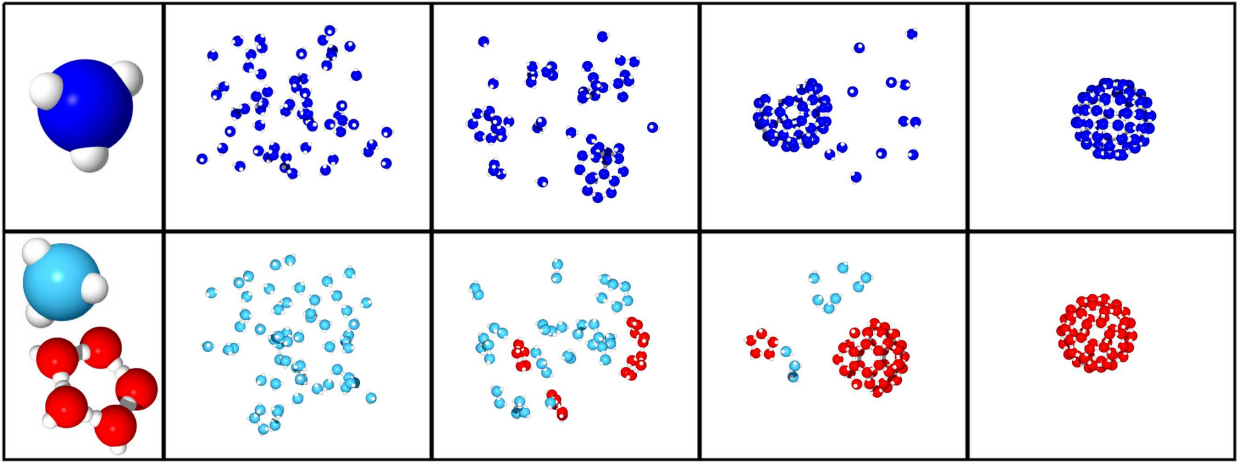}
\caption{Visualization of the computer simulations. The snapshots show the course of T1 virus capsid formation for direct (top) and hierarchical
assembly (bottom) from $n_f$=60 single monomers. In hierarchical assembly, a color change of the proteins from blue to red indicates the switch of binding characteristics upon completed formation of a pentameric capsomer.}
\end{figure}

\subsection{Direct and Hierarchical Assembly}

Direct assembly is defined as the formation of a capsid from monomers
whose bond properties remain unchanged throughout the whole
simulation.  Thus every reaction patch is active at all times. In
contrast to this unconstrained assembly mechanism, hierarchical
assembly is decomposed into multiple steps of switching of patch
reactivity depending on the configuration of the particles.  Since
pentameric and hexameric rings have been identified as key subunits in
the assembly process, we implement hierarchical assembly as switching
of reactivity upon formation of these structures. Initially only the
two patches leading to assembly of pentameric or hexameric ring
structures are active on every single protein (intra-capsomer bonds).
These patches are locked once the ring has closed, so that the
formation of these subunits is irreversible. Simultaneously to the
locking of intra-capsomer patches the binding sites which connect the
pentamers and hexamers with each other (inter-capsomer bonds) are
activated so that in a second step, formation of the capsid proceeds
via association of the capsomer rings. This collective switching in binding
properties should not be confused with the conformational switching
of individual subunits which has been used before to study structural polymorphism
\cite{Nguyen2008,Elrad2008,Nguyen2009}.

\subsection{Simulation Details}  
    
At the beginning of each simulation run the single proteins are placed
at random, non-overlapping positions in a cubic periodic boundary
box. From this configuration we let the system evolve according to the
algorithm described above with a constant timestep $\Delta t$
corresponding to a real time of $0.1$ ns. A trajectory (one simulation
run of predetermined finite length) is considered as successful if a
complete virus shell is formed within the simulation time. The
diffusive properties used here correspond to a temperature value of
T=293 K and a viscosity value of $\eta$=$2\cdot10^{-3}$Pa s. The
single proteins have a radius of $R$=1 nm and a patch radius of
$r$=0.4 nm with the center of the spherical patches placed at the
surface of the protein.  We choose the same initial concentrations for
all simulations of one virus geometry. To observe a considerable
number of association events within a reasonable time we use
relatively high concentrations of several mM.  Although these
concentration values exceed those applied in experimental setups
(several $\mu$M \cite{Schreiber2009}), this is a common practice in
simulation approaches
\cite{Nguyen2007,Rapaport2010,Schluttig2008,Schluttig2010}.  During
the simulations we record the number of clusters of size $n$, $\nu_n$
($1\leq n\leq n_f$=60 T), as well as the first passage times (FTPs) of
intermediates of specific sizes. The probability that some monomer
belongs to a cluster of size $n$ is $p(n)=(\nu_n n)/n_f$. The sum of
these probabilities is normalized to one. The average cluster size is
given by
\begin{equation}
  \overline{n}=\sum_{n=1}^{n_f}p(n) n=\sum_{n=1}^{n_f} \frac{\nu_n n}{n_f} n\ . 
\label{equation_1}
\end{equation}

\section{Results}

\subsection{Overview}

In this section we investigate the dynamics of virus capsid formation
for direct and hierarchical assembly and compare them in order to
identify their generic differences. To characterize the assembly
performance, the yield (i.e. the relative number of successful
trajectories within a given simulation time) and the first passage
times (FPTs) of selected intermediates are recorded for different
model parameters. We systematically compare both assembly mechanisms
in a parameter space ranging from $k_a$=3.0 ns$^{-1}$ to 9.0
ns$^{-1}$ and from $k_d$=$1.5 \cdot 10^{-3}$ns$^{-1}$ to $1.95
\cdot 10^{-2}$ns$^{-1}$.  For the simulations of assembly of T1
capsids we use an initial monomer concentration of $c$=4.5 mM ($60$
particles in a cubic box with side length $L$=28 nm).  Investigation
of T3 is carried out at an initial concentration of $c$=1.7 mM
($n_f$=180, $L$=55 nm).  To classify different assembly regimes we
distinguish between three different phases: the early, intermediate
and final phases which we define to be delineated by the emergence of
cluster sizes 1/3 $n_f$, 2/3 $n_f$ and $n_f$, respectively.

\begin{figure}[t]
\includegraphics[width=0.5\textwidth]{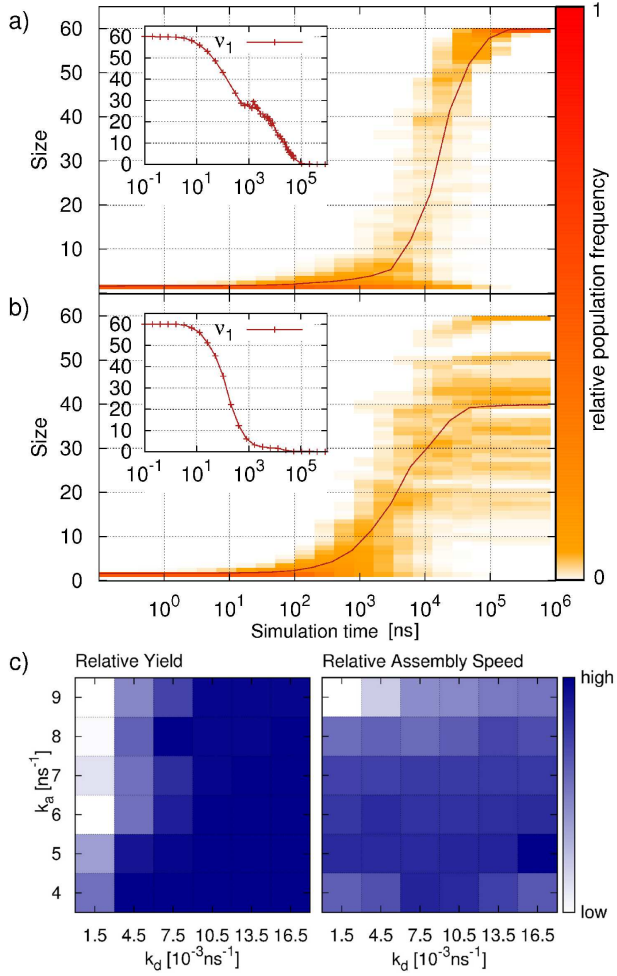}
\caption{T1 direct assembly. a) and b) show the relative population of different cluster sizes
as a function of time for a favorable ($k_a$=5.0 ns$^{-1}$, $k_d$=$13.5 \cdot 10^{-3}$ns$^{-1}$)
and a unfavorable ($k_a$=8.0 ns$^{-1}$, $k_d$=$1.5\cdot 10^{-3}$ns$^{-1}$) set of parameters, respectively.
The average cluster size is shown as solid line. In the inset the monomer population $\nu_1(t)$ is shown as a function of time.
c) Parameter space analysis of direct assembly. Relative yield (left) and relative assembly speed (right) are depicted using a heat-map representation for
various combinations of $k_a$ and $k_d$. All data are obtained from 40 independent simulation runs.}
\label{figure2}
\end{figure}

\subsection{T1 Direct Assembly}

Figures 2a and b show the temporal evolution of the relative
population of all cluster sizes for one favorable and one unfavorable
set of model parameters, respectively.  The average cluster size
$\bar{n}(t)$ (see Eq. 1) is shown as solid line and shows a sigmoidal
shape. Starting from a full set of available monomers, we observe
subsequent formation of dimers, trimers and then larger intermediate
clusters.  In the favorable case shown in Figure 2a, the distribution
always stays close to the average and complete capsid formation is
achieved. Remarkably, this successful case is also characterized by
the relatively long persistence of a monomer pool (inset to Figure
2a). The persistence of a relatively high number of monomers during
the intermediate assembly phase shows the system's capability to
reorganize and enables one dominant cluster to grow. In marked
contrast, for the unfavorable case shown in Figure 2b, the distribution of
intermediates considerably broadens. The average does not reach
complete capsid formation, and the monomer pool is depleted much
earlier. Here the intermediates are more restricted in undergoing
recombinations, many trajectories become kinetically trapped and the
average does not capture anymore the dynamics of the assembly
process. In both cases, the assembly dynamics slow down during the
final phase. This can, at least partly, be attributed to monomer
starvation as the slow-down occurs when only very few monomers are
left. The prominent features found here (sigmoidal kinetics, fast
growth after lag time, kinetic trapping, monomer starvation
in the final phase) have been found before also with coarse-grained
MD-simulations \cite{Hagan2006,Nguyen2007,Rapaport2008}.

The main difference between the two parameter sets used in Figure 2 is
that the second (unfavorable) case leads to more stable intermediates (higher $k_a$,
lower $k_d$). In Figure 2c we systematically investigate the effects of the bond
parameters on direct assembly by comparing yield and
assembly speed for different combinations of $k_a$ and $k_d$.  The
upper left corner of the parameter plots represent strong bonds (high
$k_a$, low $k_d$), while weak bonds are found in the lower right
corner (low $k_a$, high $k_d$).  The left plot shows the relative
yield averaged over an ensemble of 40 trajectories.  Direct assembly
of T1 shows a large region of high yield for dissociation rate values
above a threshold of around $k_d$=$10.5 \cdot 10^{-3}$ns$^{-1}$. Below
this value almost no successful assembly is observed.  This is due to
the limited possibilities of the intermediates to reorganize, which
results in the occurrence of kinetically trapped structures.  For low
dissociation rates we also observe a dependency of the yield on the
choice of $k_a$. In this region, lowering of the bond breaking rate
$k_d$ can at least partly be compensated by lowering of the
association rate $k_a$.

In the right plot of Figure 2c, we show the relative assembly speed as
a function of model parameters. The assembly speed is defined as the
inverse of the completion time of the capsid, $v$ =
1/FPT($n_f$). Because this quantity can be obtained only for
successful assemblies, here we average only over completed
trajectories.  In contrast to the relative yield, we see a clear
dependence of the assembly speed on the association rate $k_a$ across
the whole parameter space.  Fastest assembly is observed for
relatively low values of $k_a$. The observation that relatively high
association rates lead to slower assembly can be explained by the
increasing tendency to form more than one large cluster in the early
and intermediate phases. Thus, even for high dissociation rates, the
necessary rearrangement of the clusters slows down the assembly
process considerably. We also record a relatively high assembly speed
at low $k_d$ values where only low yield is observed. Since the
relative speed values are obtained by averaging over successful
trajectories only, these results show that, if a full capsid is
formed, it is completed within a short time.

To conclude, we see that the success of assembly in terms of yield is
mostly determined by the choice of the dissociation rate $k_d$.  For
low values of $k_d$ the system becomes kinetically trapped, while
large values of $k_d$ allow for the reorganization of the clusters.
The relative assembly speed of successful trajectories is strongly
influenced by the choice of $k_a$. Here we identify an optimum at
$k_a$=5.0 ns$^{-1}$, with speed being worse both at larger and smaller
values.  In agreement with previous studies, we observe that most
efficient assembly (i.e. high yield combined with fast capsid
completion) occurs at intermediate bond stability and that bond
reversibility is an important requirement for successful capsid
formation \cite{Rapaport2008,Hagan2011,Rapaport2010b}.

\begin{figure}
\includegraphics[width=0.5\textwidth]{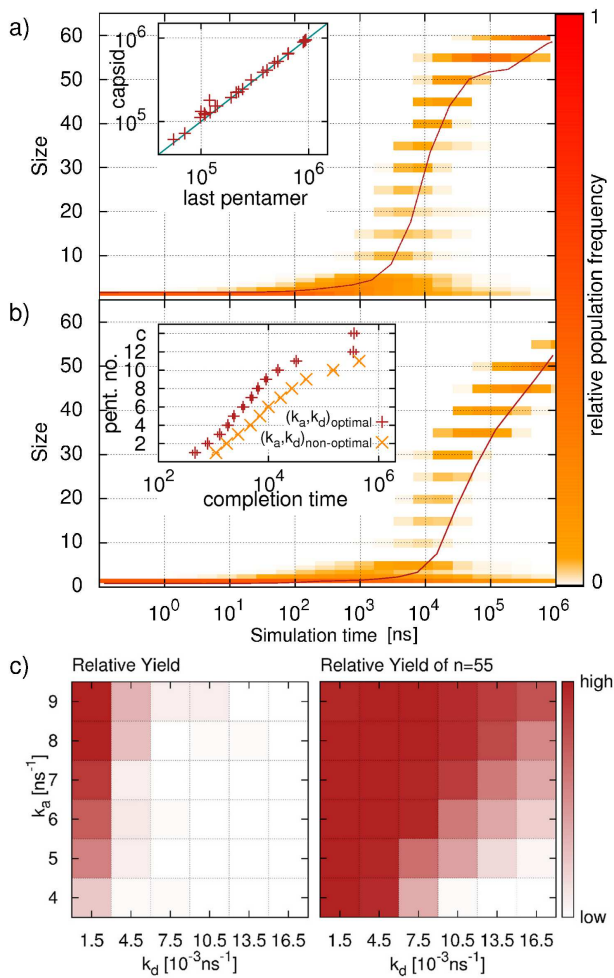}
\caption{T1 hierarchical assembly. a) and b) show the relative population for a favorable ($k_a$=8.0 ns$^{-1}$, $k_d$=$1.5 \cdot 10^{-3}$ns$^{-1}$) and unfavorable ($k_a$=5.0 ns$^{-1}$, $k_d$=$1.35\cdot 10^{-2}$ns$^{-1}$) set of parameters, respectively. The average cluster size is shown as solid line. 
In the inset of a) the FPT($n_f$) is plotted against the FPT of the last pentamer. The inset of b) shows the completion times of
the pentamers for the parameter sets analyzed in a) and b), respectively.
c) Parameter space analysis of hierarchical assembly. The relative yield of full capsids (left) and of clusters of size $n$=55 (right)
are depicted using a heat-map representation for various combinations of $k_a$ and $k_d$. All data are obtained from 40 independent simulation runs.}
\label{figure3}
\end{figure}

\subsection{T1 Hierarchical Assembly}

Hierarchical assembly of a T1 virus capsid is analyzed in a similar
manner as direct assembly.  Figures 3a and b show the evolution of
relative cluster size population and the average cluster size for
assembly under favorable and unfavorable conditions, respectively.
Due to the imposed hierarchy, clusters above pentamer size adopt only
particular size values (multiples of five).  Hierarchical assembly
under favorable conditions (Figure 3a) shows a long early phase during
which the first pentamers are formed.  The following intermediate
phase is characterized by addition of newly formed capsomers to one
dominant cluster.  A striking feature of hierarchical assembly is the
dramatic slow-down in the final phase. A majority of trajectories
remains in the $n$=55 state for a long time where all but one pentamer
have formed and joined the almost complete capsid. This can be
explained by increased monomer starvation. In hierarchical assembly,
all small clusters of sizes below five are connected by single bonds
only. Since the low monomer concentration in the final phase reduces
the frequency of diffusional encounter, it takes a long time before
the last pentameric ring can be closed irreversibly. From the inset of
Figure 3a we clearly identify the formation of the last pentamer as
the bottleneck of capsid completion in hierarchical assembly. Here the
capsid completion time is plotted against the formation time of the
last pentamer for several successful trajectories of one exemplary
parameter set. We observe that the completion of the last pentamer is
almost instantly followed by its integration into the capsid.

The assembly dynamics shown in Figure 3b for an unfavorable parameter
combination does not lead to complete assembly within the given
simulation time. In contrast to the favorable case (Figure 3a), the
association rate is lower and the dissociation rate is higher, which
results in a reduced overall bond stability. This is found to strongly
hinder the formation of the late pentamers.  Although slower, the
overall course of the assembly process is not substantially different
from the successful case in Figure 3a.  The main difference between
the two parameter combinations becomes clear by looking at the
completion times of the pentamers which are shown in the inset of
Figure 3b. We see that the pentamer FPTs of both cases follow the same
shape during the early and intermediate phases, but that for low bond
stability the completion times in the late phase are delayed. This
delay grows with ongoing assembly, so that the final pentamer does not
close within the simulation time. Here the negative effect of low
monomer concentration on capsomer assembly, which was discussed
earlier, is amplified by the low bond stability.
  
To quantify the effects of different combinations of $k_a$ and $k_d$
on hierarchical assembly of T1, we again perform a systematic
investigation of the bond parameter space as shown in Figure 3c.
Considering the relative yield of complete capsids (Figure 3c, left
image), we observe that only a narrow range of parameters leads to a
considerable fraction of successful trajectories. High yield is only
observed at high bond stabilities (high $k_a$, low $k_d$) in the upper
left corner of the parameter plot. To take into account the critical
role of the formation of the last pentamer in our simulations, we also
show the yield of almost finished capsids ($n$=55) at the end of the
simulation time (Figure 3c, right image). The region where we observe
almost finished capsid is considerably expanded and a large fraction
of trajectories reaches $n$=55 in the upper left corner of parameter
space. The yield decreases along the diagonal from high towards low
bond stability values (lower right corner). It becomes clear that the
unfavorable parameter combinations do not show kinetically trapped
states as they occur in direct assembly, and that most trajectories
are close to capsid completion. The high yield of almost finished
capsids and the lack of trapped trajectories suggests that the bond
hierarchy promotes successful capsid completion, but is vulnerable to
monomer starvation.

\subsection{T1 Direct versus Hierarchical Assembly}

\begin{figure}
\centering
\includegraphics[width=\textwidth]{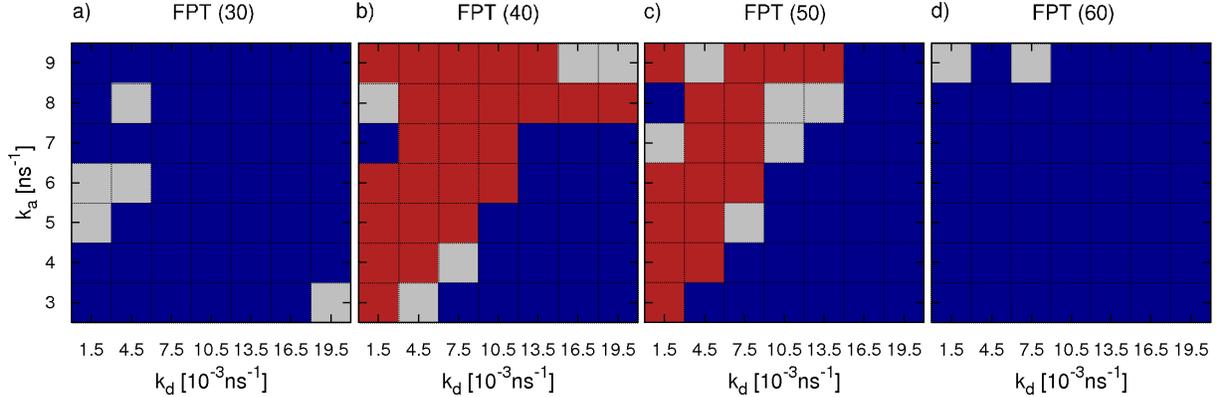}
\caption{Comparison of T1 direct and hierarchical assembly. We evaluate a) FPT(30), b) FPT(40), c) FPT(50) and d) FPT(60) for different parameter combinations ($k_a$, $k_d$). Blue fields indicate points at which the respective FPT for direct assembly is smallest while red fields identify hierarchical assembly to be faster. Points where no clear distinction is possible are colored in gray. Every data point is obtained from 40 simulation runs.}
\label{figure_4} 
\end{figure}

The above analysis has revealed a marked difference between direct and
hierarchical assembly schemes.  From Figures 2 and 3, it is also clear that
the final state of a trajectory is strongly affected by the finite
length of the simulation and provides only limited information on the
dynamics of assembly. In particular hierarchical assembly depends
strongly on the formation of the last pentamer and suffers from
monomer starvation in the final phase.  To evaluate in more detail the
performance of the assembly process in its different phases, we
systematically compare the FPTs of certain intermediates for both
direct and hierarchical assembly.  The results are depicted in Figure
4 in a sequence of phase diagrams. Blue areas are those where direct
assembly performs better while red indicates parameter combinations
where hierarchical assembly is faster. Points where a clear
distinction is not possible are shown in gray (difference of direct and hierarchical FPTs less than $10\%$ of the sum of both FPTs).

For the first emergence of intermediates of half the capsid size (FPT(30), Figure 4a), direct assembly is faster throughout the whole parameter space. This is related to the earlier observation of an extended initial phase of hierarchical assembly when compared to direct assembly (see Figure 3). It can be explained by the fact that the monomers in direct assembly exhibit three active binding sites and thus easily form clusters of considerable size.
Since hierarchically assembling monomers are designed to form flat pentamer rings, only the two patches forming intra-capsomer bonds are active until full capsomers are formed. Thus the number of fruitful encounters is reduced remarkably, which leads to the observed slow-down of the initial phase.  

Looking at the FPTs for the two-third assembled capsid (FPT(40), Figure 4b), we see a large region in the upper left part of the parameter space (high $k_a$, low $k_d$) where hierarchical assembly
is now able to overtake direct assembly. This can be attributed to two effects. Firstly, hierarchical assembly speeds up once a pool of capsomers is available. Secondly, direct assembly is slowed down at high bond stabilities. Since the combination of fast formation of large, stable clusters in the early phase (due to high $k_a$) and slow dissociation of small clusters leads to a small number of free monomers, the dominant cluster grows only slowly. In the region of lower bond stability, direct assembly remains faster. Here the increased ability of un- and rebinding of single proteins allows for fast rearrangement, leading to a sufficiently large supply of free monomers so that the dominant cluster can easily grow beyond $n$=40. Simultaneously the pentamer rings in the hierarchical setup form slower than at high bond stabilities.
   
The difference between the two assembly mechanisms becomes even more evident when looking at FPT(50) (Figure 4c). 
At low $k_d$ values, direct assembly experiences kinetic trapping. As a consequence, hierarchical assembly is superior
for almost all small $k_d$ values, also at points where the question of dominance remained undecided for FPT(40).
The parameter region of weak bonds where direct assembly is faster than hierarchical one is observed to extend during the step from FPT(40) to FPT(50) 
(lower right corner of parameter space). Under these conditions the effect of beginning monomer starvation delays the pentamer completion of hierarchical assembly.

For the assembly speed of the complete virus capsid (FPT(60), Figure 4d), direct assembly dominates again across almost the whole parameter space.
Only at very high bond stabilities hierarchical assembly shows lower or comparable FPT values. This is not surprising taking into account the results
for the overall yield of hierarchical assembly (Figure 3c) and underlines the large impact of monomer starvation on hierarchical assembly.

We conclude that hierarchical assembly is not always better than direct assembly. Direct assembly performs better
both at the initial and final phases. During the intermediate phase,
however, hierarchical assembly is more successful, because it does not suffer from stable bonds preventing structural rearrangements.
Due to the limited number of possible interactions, hierarchical assembly is unlikely to get trapped in sub-pentameric units.
In general, for hierarchical assembly parameter combinations resulting in high bond stability are favorable. At these values we observe kinetic
trapping of most of the directly assembling systems. In addition, the symmetry of the pentamers themselves and the low complexity of their
interactions prevent them from getting trapped in large clusters incompatible with the final
capsid. For T1, this favors the step-wise build-up of the target structure.

\begin{figure}
\includegraphics[width=0.5\textwidth]{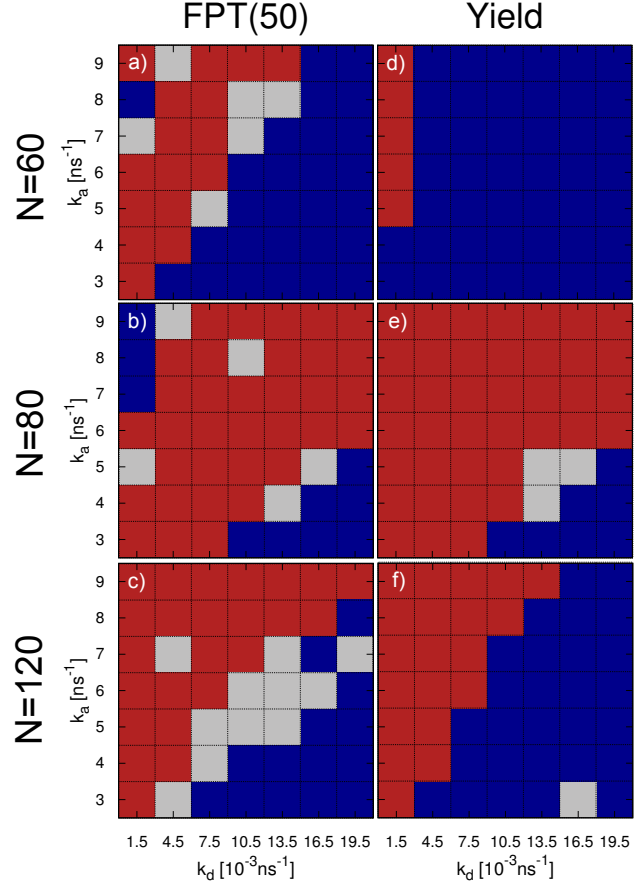}
\caption{Effect of initial number of monomers on T1 assembly. Comparison of FPT(50) (a-c) and yield (d-f) for direct and hierarchical assembly with an initial number of N=60, N=80 and N=120 monomers, respectively. Blue fields indicate points at which the FPT for direct assembly is smallest or the yield is largest while red fields identify hierarchical assembly to be faster or the yield to be higher. Points where no clear distinction was possible are colored in gray. Every data point is obtained from 45 simulation runs.}
\end{figure}

\subsection{T1 Effect of Initial Number of Monomers}

Until now we have used exactly as many monomers as needed to form one
complete capsid. In experiments, monomers are likely to be present in
surplus or to be provided with a certain rate. To study the effect of
the limited number of monomers on our simulation, we next increase the
initial supply to N=80 and N=120 monomers while keeping the
concentration constant by enlarging the simulation box. For the case
N=80, a surplus of 20 monomers will be present upon formation of a
complete capsid. For the case N=120, two capsids might be formed in
parallel and thus the benefit of an increased initial monomer
concentration might be shared by them in a complex manner.  Figures 5a-5c
show a comparison between direct and hierarchical assembly of the FPT(50) for an initial number of N=60, N=80
and N=120 monomers, respectively.  As in Figure 4, blue fields indicate
that direct assembly has a lower FPT(50), red fields mark parameter
pairs for which hierarchical assembly is faster and for gray fields no
clear distinction is possible. We see that the comparison of both
mechanisms leads to similar results for all setups. When increasing
number of initial monomers we observe a slightly larger region of the
parameter space in which hierarchical assembly becomes favorable. This
is not surprising, as we identified monomer starvation to strongly
hinder the final capsid completion for hierarchical assembly. However,
in general the effect of monomer starvation seems to have relatively
little impact on the relative efficiency of the two different assembly
schemes for clusters of size N=50.

Figures 5d-5f show the yield of the first capsid within simulation time
for an initial number of N=60, N=80 and N=120 monomers,
respectively. Here again red indicates a higher yield of hierarchical
assembly while blue indicates a higher yield of direct assembly. Gray
marks parameter pairs with the same yield. In contrast to the FPT(50),
we can clearly see that increasing the initial number of monomers
results in a largely expanded parameter space in which hierarchical
assembly is favorable. This shows that monomer starvation affects the
final phase of hierarchical assembly in particular as it has been
inferred in the previous section. In fact hierarchical assembly
performs well throughout the whole parameter space and shows high
yield for intermediate and weak bonds. At very high bond strength we
even observe some trapping for hierarchical assembly. However, direct
assembly still strongly suffers from kinetic trapping so that the
parameter space corresponding to high bond strength remains clearly
dominated by hierarchical assembly. We also note that increasing the initial 
number of monomers from N=80 to N=120 does not lead to a further promotion of 
hierarchical assembly, presumably because now two capsids form
in parallel, each drawing monomers in a similar manner as before
for N=60.

To conclude, we find that our main results from the previous section
remain valid for an increased number of initial monomers. Hierarchical
assembly is favorable at high bond strength due to the decreased
possibility of trapping while direct assembly is favorable at low bond
strength allowing for fast reorganization of large clusters. In
general, we expect that our results also carry over to even larger
systems.

\begin{figure}
\includegraphics[width=0.7\textwidth]{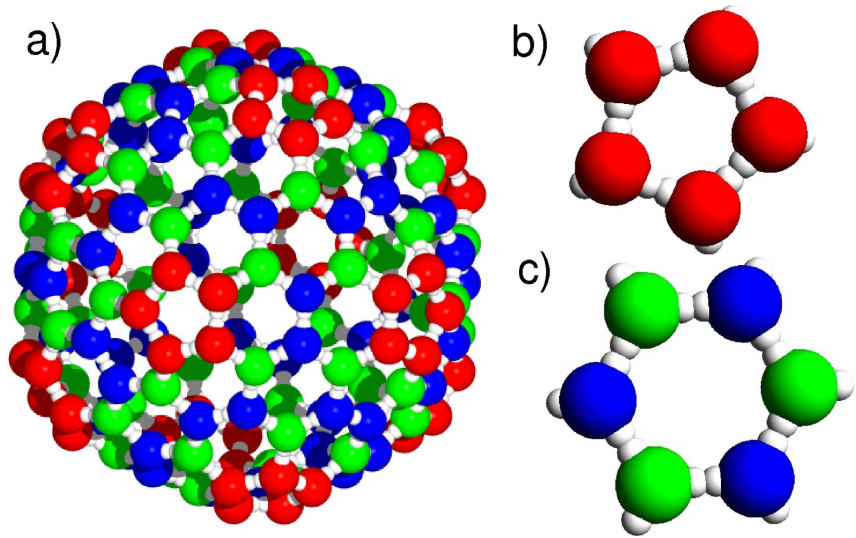}
\caption{Model for T3 virus capsid. (a) Visualization of the T3 virus capsid and its capsomers of (b) pentameric and (c) hexameric structure. The hexamer is composed of two different protein types.}
\end{figure}
  
\subsection{T3 Direct versus Hierarchical Assembly}

Given the results for T1 virus assembly, we now ask how they carry
over to more complicated geometries like T3 viruses.  In this section
we compare the characteristics of direct and hierarchical assembly of
T3 viruses which are composed of $n_f$=180 monomers.  Now we place
again exactly the number of monomers needed for the formation of one
complete capsid into the simulation box.  While in the hierarchical
assembly of T1 viruses the capsid was built from pentameric subunits
only, T3 virus capsids consist of 12 pentameric and 20 hexameric
capsomers. Figure 6 shows a model capsid which, in the hierarchical
case, assembles from two different subunits. While the pentamers are
formed from identical proteins, the hexamers contain two different
particle types.  Due to the increased complexity of the T3 capsid, we
observe only a small range of bond parameters to lead to high yield
for direct assembly in our computer simulations.  Moreover, we are not
able to identify a parameter combination that allows successful
hierarchical assembly within the used simulation time. This is caused
by the lowered concentration of individual species of monomers which
leads to a dramatic slow-down of capsomer formation in the final
phase. As hierarchical assembly reaches the largest cluster sizes at a
high association rate of $k_a$=9.0 ns$^{-1}$, we now systematically
analyze the effect of different dissociation rates $k_d$ ($5 \cdot
10^{-4}$ns$^{-1} \leq k_d \leq 1.35 \cdot 10^{-2}$ns$^{-1}$) while
keeping $k_a$ fixed.

\begin{figure}
\includegraphics[width=0.5\textwidth]{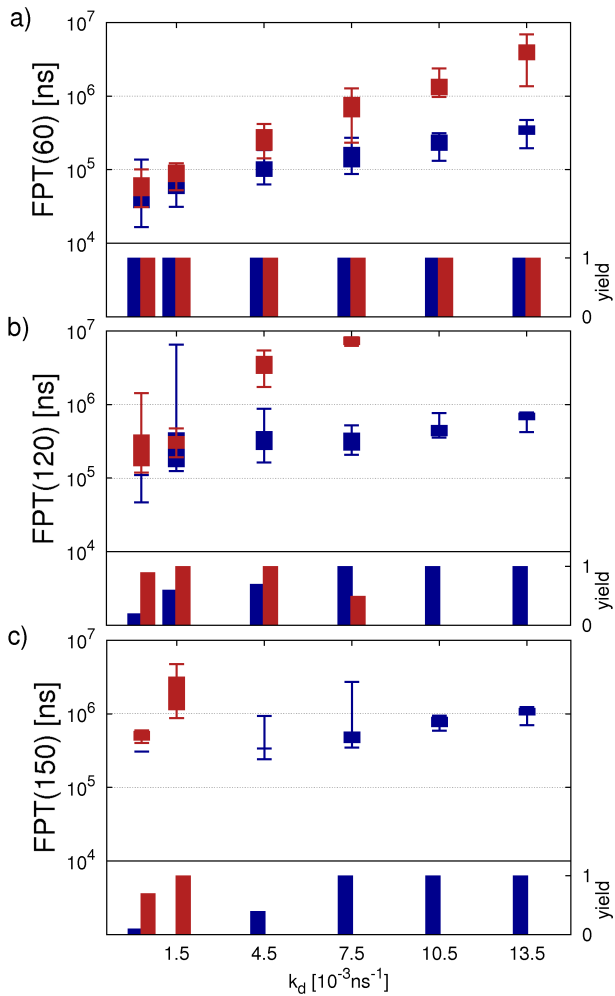}
\caption{Comparison of T3 direct and hierarchical assembly. a), b) and c) show the first passage times FPT(60), FPT(120) and FPT(150) together with the relative yield of the corresponding cluster size for fixed $k_a$=9.0 ns$^{-1}$. Blue and red boxes show the results for direct and hierarchical assembly, respectively. The data points are obtained from 10 simulation runs. The maximum simulation length of $9\cdot10^{6}$ns represents the upper boundary of the FPT values.}
\end{figure}

Figure 7 shows different FPTs ($1/3$ $n_f$, $2/3$ $n_f$ and $5/6$ $n_f$) for direct (blue color) and hierarchical (red) assembly for $k_a$=9.0 ns$^{-1}$ and varying dissociation rate.
The FPTs are complemented with yield histograms showing the relative number of trajectories which reached the corresponding size within the simulation time. 
From Figure 7a we immediately see that all trajectories in the investigated parameter interval have grown beyond a cluster size of $n$=60 at the end of the simulation.
Comparison of the FPTs for direct and hierarchical assembly reveals assembly speeds of the same magnitude at low values of $k_d$. 
With growing dissociation rate the FPTs increase for direct as well as for hierarchical assembly. This is not surprising since a lower bond stability leads to an increased number of dissociation events and a slower cluster growth. 
The FPTs of direct assembly increase only moderately (about one order of magnitude) compared to those of hierarchical assembly (two orders of magnitude). 
This extreme sensitivity of hierarchical assembly is caused by the strong impact of the low bond stability on capsomers formation.
The effect was already observed in hierarchical assembly of T1 and is amplified here due to the presence of several protein types and the resulting lowered effective initial concentration:
The number of fruitful monomer encounters is not only reduced by the smaller number of active patches compared to direct assembly, but also by the limited number of suitable binding partners. 
As a consequence of the dramatic slow-down of hierarchical assembly with increasing $k_d$, we observe zero yield of intermediates of size $n$=120 above a threshold around
$k_d$=$7.5\cdot 10^{-3}$ns$^{-1}$ (Figure 7b). On the contrary, we record a decrease in the yield of direct assembly below this $k_d$ value. 
This can be explained with the occurrence of kinetic trapping which we already encountered in T1 direct assembly. 
Analysis of the corresponding FPT values of so far successful trajectories reveals that, despite the trapping tendency, the speed of direct assembly is still comparable to that of hierarchical assembly at low $k_d$ values.
For even larger cluster sizes (FPT(150), Figure 7c) we see further partitioning of the parameter space. Above a threshold around $k_d$=$4.5 \cdot 10^{-3}$ns$^{-1}$,
no hierarchical assembly is observed, while below this value, only one directly assembling trajectory reaches this size. 

\subsection{T3 Effect of Initial Number of Monomers}

\begin{figure}
\includegraphics[width=\textwidth]{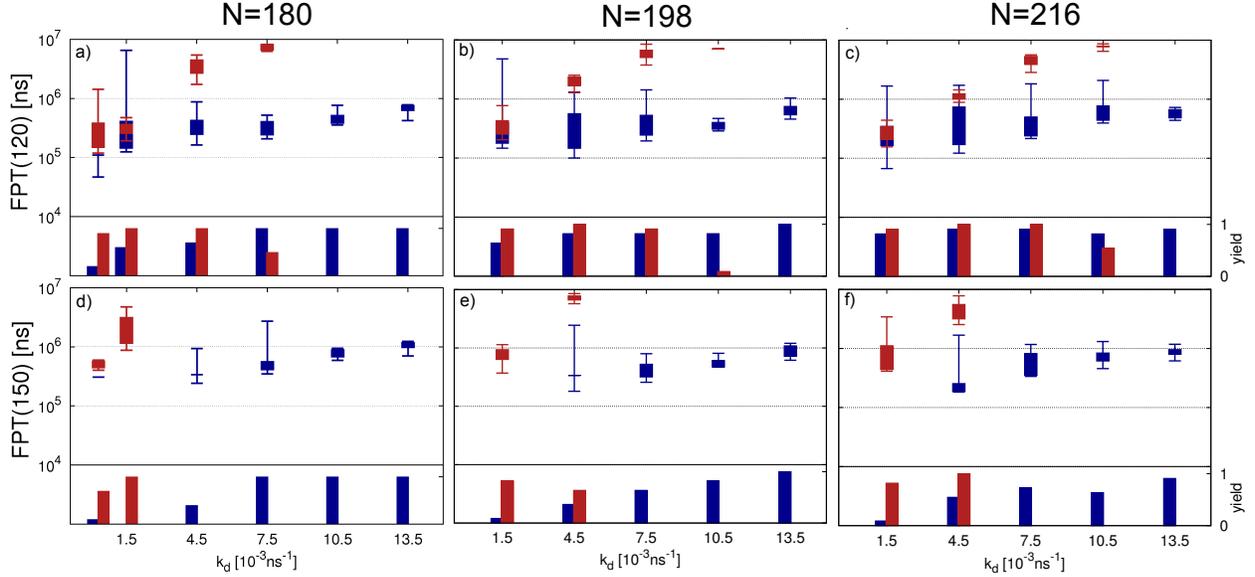}
\caption{Effect of initial number of monomers on T3 assembly. Comparison of T3 direct and hierarchical assembly for an initial number of N=180, N=198 and N=216 monomers. a)-c) show the FPT(120)  and d)-f) show the FPT(150) together with the relative yield of the corresponding cluster size for the respective initial number of monomers while. Blue and red boxes show the results for direct and hierarchical assembly at a fixed $k_a$=9.0ns${-1}$ in a range of $k_d=1.5\cdot 10^{-3}$ns$^{-1}-13.5\cdot 10^{-3}$ns$^{-1}$.For N=180 a additional value at $k_d=0.5\cdot10^{-3}$ns${-1}$is shown. The data points were obtained from at least 10 simulation runs with a maximum length of $9\cdot10^{6}ns$}
\end{figure}

As for the assembly of T1 capsids, we again investigate the role of an
increased initial number of monomers on the simulation results for the
T3 capsid. We increase the initial number of monomers by $10\%$ and
$20\%$ (without changing the concentration) and record the FPTs for
these simulations. In Figures 8a-8c the FPT(120) and the yield of
clusters of size 120 is shown for an initial number of N=180, N=196
and N=216 monomers, respectively. As in the previous section we
explore the effect of varying $k_d$ while keeping $k_a=9.0$ns$^{-1}$
fixed. Comparing the FPT(120) for the different setups we see that
above $k_d=1.5$, hierarchical assembly becomes faster for an increased
initial number of monomers. Direct assembly in contrast is only
slightly affected throughout the parameter space. When looking at the
yield of clusters of size 120 within simulation time ($9\cdot
10^{6}$ns), we clearly see the positive effect of an increased initial
number of monomers on hierarchical assembly for weaker bonds (higher
$k_d$). However, it remains worse than direct assembly at these bond
strengths. These findings are in agreement with the effect observed
for T1 when increasing the initial number of monomers. While the
dynamic of direct assembly is only weakly affected by the initial
number of monomers, hierarchical assembly suffers less from the effect
of monomer starvation at weak bond strength. Considering the FPT(150)
we again see a complete separation of the parameter space into one
region in which only direct assembly is observed and another region in
which hierarchical assembly dominates. Looking at the yield we see
that for an initial number of 180 monomers hierarchical assembly is
only observed for $k_d\leq1.5\cdot 10^{-3}$ns while this region
expands to $k_d\leq 4.5\cdot 10^{-3}$ns for an increased initial
number of monomers. It might be possible that the parameter space in
which hierarchical assembly is favorable expands further for a larger
increase of the initial number of monomers, similar as it was observed
for T1 (Figure 5). However, it seems that the favorable effect of an
increased initial number of monomers is weaker for T3 capsids than for
T1 capsids due to the more complex geometry. In the following section
we will investigate the role of complexity of the T3 capsid for the
hierarchical assembly of a T3 capsid.

\subsection{Capsomer Formation in T3 Hierarchical Assembly}

In order to further investigate the effects that slow down
hierarchical assembly, we now analyze the dynamics of hexamer and
pentamer formation both with computer simulations and a master
equation approach.  To compare the FPTs for pentamer and hexamer
formation, we scale these values with the number of monomers per
capsomer ring. This linear scaling is based on the assumption that the
mean time for a net addition of monomers to small ring-forming
clusters is independent of the cluster size.  This simplification in
particular neglects the increased number of decay paths of hexamers
compared to pentamers.  However, the assumption seems justified for
the present case of high bond stabilities (high $k_a$, low $k_d$), at
least for the early and intermediate phase of assembly.

\begin{figure}
\includegraphics[width=\textwidth]{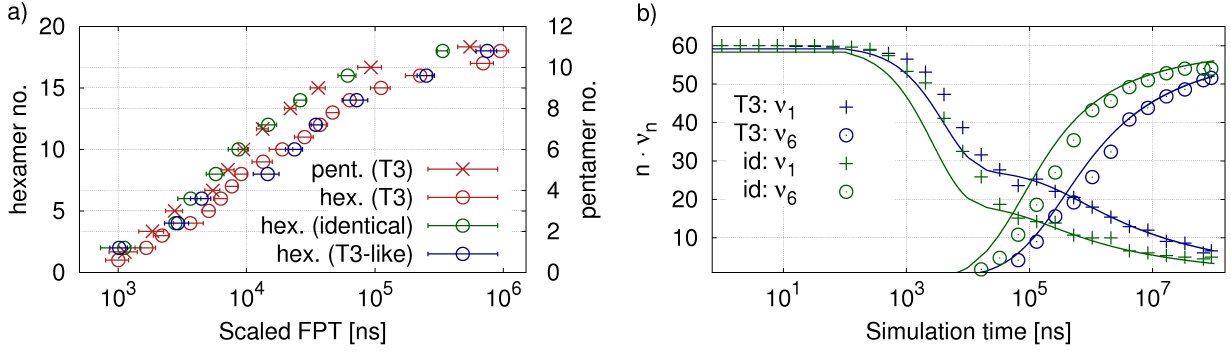}
\caption{Analysis of T3 capsomer assembly. a) FPTs of pentamer and hexamer capsomers emerging during T3 hierarchical assembly are compared to those of T3-like and identical hexamers from the down-scaled simulations. All simulations use $k_a$=9.0 ns$^{-1}$ and $k_d$=$1.5\cdot 10^{-3}$ns$^{-1}$. As there are different total numbers of capsomers to be formed of each type, we compare the relative progress of assembly by plotting all species in the same plot with different scales (1 to 12 for pentamers, 1 to 20 for hexamers). FPTs are scaled with the ring size. All values are obtained from 10 independent simulation runs each. b) Relative cluster size population $n\cdot\nu_n(t)$ during hexamer assembly in the down-scaled systems. The results for monomers ($\nu_1$) and complete capsomers ($\nu_6$) from the analytical master equation approach (lines) are compared to the simulation data (symbols) for T3-like and identical hexamers.}
\end{figure}

In Figure 9a the average capsomer formation times from T3 assembly at
the most promising parameters identified from Figure 7 are shown (now
again for N=120).  We find that during T3 capsid assembly hexamers
form slower than pentamers (for the same sequential number).  The
difference between the completion times increases with time (the last
hexamer data point, no. 18, is an exception to this rule since its FPT
is artificially cut down to lower values by the finite length of the
simulation).  In order to investigate whether this is caused by the
different relative densities of monomers forming pentamers (60/180)
and hexamers (120/180) or a result of the increased complexity of the
hexamer rings, we perform a separate set of simulations. In this
complementary simulation we compare the assembly of hexamers
consisting of one type of protein (identical hexamers) and hexamers
built from two different types of proteins (T3-like hexamers). To
reduce the computational effort we downscale the system to half its
size while preserving the concentration (i.e. assembly of 10 hexamers
in the presence of 30 pentamer-forming monomers).  In Figure 9a the
hexamer-FPTs from the complementary simulation are compared to those
at the same relative positions in the assembly process of the full
simulations. The FPTs are again scaled with the ring size.  While the
dynamics of the identical hexamers follow the course of the pentamers
in the full simulation, the FPTs of the T3-like hexamers and the T3
hexamers of the full simulation are in good agreement.  This
observation suggests that the delay in hexamer formation observed in
the full simulation is caused by the two-type complexity of the
hexamers compared to the uniformly structured pentamers.

To complement this investigation we use an analytical master equation approach to perform a closer analysis of the dynamics of parallel assembly of several hexamers.
Here we develop a set of equations which gives analytic results for the number of clusters of size $n$, 
$\nu_n(t)$ ($1\leq n \leq n_f$=6), as a function of association and dissociation rate.
The time evolution of the macroscopic quantity $\nu_{n}$ is the result of reactions between clusters of all sizes $k$ 
which cause a change of $\nu_{n}$.  
We introduce the association rate $a$ for successful binding of two clusters per unit time and the dissociation rate $b_{nk}$ which denotes the rate for decay of a cluster of size $n$ to two daughters of sizes $k$ and $(n-k)$. 
$b_{nk}$ is composed of the dissociation rate per bond per unit time, $b$, and a factor $d_{nk}$ which quantifies the probability for the 
decay of a cluster of size $n$ to a constellation where one of the daughters is of size $k$. 
$d_{nk}$ is determined by the ratio of total dissociation probability (proportional to the number of bonds which compose $n$) 
and the probability of the decay products to have the required size. 
The population $\nu_i$ increases by the decay of clusters with sizes larger than $i$, so that $d_{ji}$ ($i<j<n_f$) is always positive. For these cases we find $d_{ji}$=2 for each pair $j$, $i$, since the decay from $2i$ to two daughters of sizes $i$ accounts for a double increase of $\nu_i$. The factor $d_{nn}$ denotes the total decay probability of a 
cluster, it is thus negative and proportional to the cluster size. Here we use $d_{nn}=-(n-1)$ for every $n<n_f$.
We account for one-step processes only, which means we focus on transitions where two clusters merge or one cluster falls apart 
into two daughter clusters. 
If we assume that the formation of the complete hexamer ring is irreversible and that the total number of particles $N$ is preserved, the 
complete set of equations describing the time evolution of the system reads 

\begin{align}
&\dot{\nu}_{n}(t)=\underbrace{ \sum_{k+l=n} a \nu_{k}(t)\nu_{l}(t) }_{\substack{\textnormal{growth by association}\\\textnormal{of smaller clusters}}}
- \underbrace{ \nu_{n}(t) \sum_{k=1}^{n_f-n} a \nu_{k}(t) }_{\substack{\textnormal{decrease by association}\\\textnormal{with other clusters}}}
+\underbrace{ \sum_{k=n}^{n_f-1} b_{kn} \nu_{k}(t) }_{\substack{\textnormal{growth/decrease}\\ \textnormal{by dissociation events}}} \\
 &\dot{\nu}_{n_f}(t)=\sum_{k=1}^{n_f / 2} a \nu_{k}(t)\nu_{(n_f-k)}(t) \hspace{17.5mm} \textnormal{(boundary condition)} \\
 &\sum_{n=1}^{n_f} n \cdot \nu_{n}(t) = N \hspace{35mm} \textnormal{(constraint)} 
\end{align}

Numerical evaluation with the initial condition $\nu_1(t$=0)=$N$ gives the time evolution of all cluster size populations $\nu_n$. 
By fitting the set of equations to the course of all $\nu_n(t)$ from the complementary simulation ($n_f$=6, $N$=60), 
we obtain parameter combinations $a$, $b$ which reproduce the observed assembly dynamics. 
Under the constraint that the dissociation rate $b$ per bond 
is constant for identical and T3-like hexamers (since the simulations apply the same $k_d$), we find the following parameters: 
$a^{id}$=$2.4\cdot 10^{-6}$ns$^{-1}$ (identical hexamers), $a^{T3}$=$1.2\cdot 10^{-6}$ns$^{-1}$ (T3-like hexamers) 
and $b$=$9\cdot 10^{-5}$ns$^{-1}$.  
In general, all $\nu_n(t)$ are reproduced well. This suggests that the assumption of a constant association rate $a$ per bond,
independent of the sizes of the encountering clusters, is a reasonable approximation for the formation of small rings. 
The results for $\nu_1(t)$ and $\nu_6(t)$ are displayed in Figure 9b together with the simulation data points for both types of hexamers.
The early phase is the region which exhibits the largest discrepancies between data and ME results, while the final phase of assembly shows a high level of consistency. This can be explained by the fact that the rate equation framework does not include any spatial constraints and is thus not able to reproduce the same sort of lag time before the first protein reactions as was observed in the simulations, where the randomly distributed particles react only after diffusional mixing leads to the first encounter events. This is also the reason why the difference between the cases of T3-like and identical hexamers becomes visible in the simulation data only after a certain time, while the ME results differ from the very first iteration step (see Figure 9b). 
Since the rate equations do not contain a diffusional component, the coefficients $a$ and $b$ can not be directly related to the 
simulation parameters $k_a$ and $k_d$. While $k_a$ determines the rate of transition from encounter to a bound state,
$a$ as well includes the formation of diffusional patch overlap. 
Their relation is defined as $a=k_a/(N_A \cdot V)$, where $V$ is the simulation box volume. Using the initial concentration
$c=N/(N_A \cdot V)$, we find the expression $a = k_a \cdot \frac{c}{N}$.
$a$ is thus, as expected, proportional to the initial 
monomer concentration in the simulation box. Applying this relation to the fit parameters using the effective initial concentration of the protein types, we estimate the overall association rate values to be $k_{a^id}^{fit}$=$3.3\cdot10^7$s$^{-1}$M$^{-1}$ and $k_{a^T3}^{fit}=6.5\cdot10^7$s$^{-1}$M$^{-1}$. 
The fact that the association rate $a^{id}$ for identical hexamers is about twice the value found for $a^{T3}$ 
confirms that the difference in assembly dynamics for identical and T3-like hexamers has its origin in a reduced
association rate, caused by reduced encounter of matching protein types. 
Our observations suggest that the association rate decreases linearly with increasing number of bond partners in the system and 
thus the number of different protein types needed to form a capsomer ring. 
When comparing our values for the diffusional encounter rate to data from experiments, we see that we overestimate the association rate. In general, the association rate for bimolecular binding reactions is experimentally found to lie between $4\cdot10^6$ and $10^7$s$^{-1}$M$^{-1}$ \cite{Schreiber2009}. Absence of long-ranged forces, as it is the case for our simulation framework, is predicted to push the rates below $10^6$s$^{-1}$M$^{-1}$ \cite{Northrup1992}. 
The reason for our relatively high estimates for the encounter rate could be the treatment of dissociation as a stochastic event without
immediate relocation of the partners. In the present implementation, two patches stay in an encounter after dissociation and their
movement is subject to the cluster mobility.
We assume this to cause an overestimation of rebinding frequencies which results in an increased association constant.
Whereas the association rate constant can be related to other results, there is no such argument for the value of $b$.

\section{Conclusion}

Understanding the biophysical principles underlying the self-assembly
of virus capsids is of fundamental importance for biology and
medicine, and might also promote novel applications in materials
science. Here we have presented a Brownian dynamics study of the
assembly of icosahedral virus capsids. Using a patchy particle model
without potentials, our simulations are relatively fast and therefore
we are able to obtain good statistics with relatively modest computing times. 
One special strength of our approach is the rigorous treatment of translational
and rotational diffusion, with the motility matrices for any cluster
shape calculated on the fly. Our approach is particularly suited to
focus on the effect of a bonding hierarchy on the performance of the
assembly process.  The hierarchy was established by an event-driven
switching of bond characteristics upon the formation of capsomer
rings, which have earlier been identified as key intermediate
structures of the assembly pathway of some icosahedral viruses
\cite{Johnson1997,Tonegawa1970,Salunke1986,Flasinski1997,Willits2003,Hanslip2006,Oppenheim2008}.
We first conducted a detailed comparison of direct versus hierarchical
assembly for T1 viruses.  To elucidate the effects of an increased
complexity of the capsid geometry on the formation of the capsomer
rings, we then performed a detailed analysis of capsomer assembly for
T3 viruses, including a master equation approach complementing the
computer simulations.

Our results for direct assembly of T1 virus capsids show that capsid
completion is only successful if the bonds are weak enough to allow
for a sufficient number of unbinding and reorganization
events. Otherwise kinetically trapped clusters appear. These findings
are in good agreement with the results of previous approaches
\cite{Hagan2006,Rapaport2008,Hagan2011,Rapaport2010b}.  In marked
contrast, hierarchical assembly performs better for high bond
stabilities, as the imposed hierarchy reduces kinetic trapping.
However, hierarchical assembly is more vulnerable to monomer starvation in the final 
phase. This effect has previously been observed in other
approaches for direct assembly \cite{Hagan2006,Mukherjee2010}, but it is even more severe
for hierarchical assembly specifically studied here. 
Comparison of  direct and hierarchical assembly at various
reveals that hierarchical assembly, although slower in the early phases, is able to outperform direct 
assembly at high and intermediate bond strength. 
This is due to the fact that capsids assembling from highly symmetric capsomers do not require fundamental
reorganizations to achieve large cluster size, as it is the case in direct assembly. 

The analysis of T3 virus assembly shows that the effects apparent for
T1 viruses become amplified by the increased complexity of the capsid
geometry. In general, the assembly process of T3 viruses is slower due
to the size of the capsid and the increased complexity of the protein
interactions. Starting with exactly 180 monomers the parameter space
for successful direct assembly is narrowed and we do not observe any
complete capsids in hierarchical assembly within the used simulation
times.  Investigation of the course of assembly of the two mechanisms
reveals that they both perform best in distinct regions of the
parameter space. Increasing the initial number of monomers we find
that hierarchical assembly performs better while direct assembly
remains widely unaffected. However, we still observe that both
mechanisms are favorable in distinct regions of the parameter space.
To analyze the effect of geometric complexity on capsomer formation
during hierarchical assembly, we perform a closer analysis of assembly
of different capsomer types.  The results show a significant slow-down
of capsomer formation with increasing structural complexity, which
explains why we do not observe any full T3 virus capsids in the
hierarchical setup within the given simulation time.  These findings
suggest a further slow-down for the assembly dynamics of more complex
capsids such as T4 and T7 for hierarchical assembly.

Computer simulations of virus assembly are usually carried out with a
fixed number of initial monomers and therefore necessarily lead to
monomer starvation in the final phase. \textit{In vivo}, this
constraint should be less relevant than in our simulations. Once a
cell is infected by a virus, one expects to see a constant production
rate for viral proteins, and therefore monomer starvation should be
less of an issue. It would be interesting to test if in such a
situation, hierarchical assembly becomes even more favorable than
found here.  In computer simulations, this could be done by
continuously adding new monomers and removing completed capsids. We
leave this to future studies as it would entail to introduce at least
two more model parameters, namely the rates for monomer injection and
capsid removal, as well as explicit rules on the spatial positioning
of the new monomers.  A similar study could be done experimentally for
viruses which self-assemble \textit{in vitro}, although here too there
might be technical problems to implement such procedures. For
\textit{in vivo} systems, such studies would depend very much on the
details of the virus assembly of interest, in particular on the
spatial coordination in regard to the different cellular compartments.

Our simulation framework has great potential for further investigation
of assembly of icosahedral viruses, i.e. capsids with higher T-number
(T4, T7,...). Although larger simulation times become necessary, they
are potentially much smaller than the ones required for less
coarse-grained approaches, including patchy particle models with
interaction potentials or coarse-grained molecular dynamics
simulations. A particular strength of our approach is the possibility
to switch patch reactivity during the assembly process. This suggests
to investigate even more complex ways to build virus capsids. Our
approach could also be applied to other interesting cases of protein
assembly, for example to the actin cytoskeleton, for which different
regulatory proteins lead to changes in local reactivity.
  
We conclude that it might be beneficial for icosahedral viruses to
assemble hierarchically, since it prevents kinetic trapping and allows
for faster formation of larger structures. Our results suggest that
hierarchical assembly performs better than direct assembly for high
and intermediate bond stability, while direct assembly is favorable
for weak bonds allowing for fast reorganization. For complex viruses,
our study suggests that the problem of monomer starvation and critical
concentrations has to be addressed for each type of monomer
separately, thus making the process more vulnerable for fluctuations
in the supply chain and imposing limits to the overall degree of
complexity. The partitioning of parameter space into favorable regions
for direct versus hierarchical schemes becomes even stronger for more
complex capsid geometries and suggests ways to design optimal assembly
schemes for different molecular species.

\begin{acknowledgments}
We thank Jakob Schluttig for providing the initial computer code
and for helpful discussions. HCRK was supported by fellowships from
the Landesgraduiertenstiftung Baden-Wuerttemberg and from
the Cusanuswerk. USS is member of the CellNetworks
cluster of excellence at Heidelberg.
\end{acknowledgments}

\end{document}